# Dynamic 3-D measurement based on fringe-to-fringe transformation using deep learning

HAOTIAN YU,[1,2,3] XIAOYU CHEN,[1,2,3] ZHAO ZHANG,[1,2] YI ZHANG,[1,2] DONGLIANG ZHENG,[1,4] AND JING HAN,[1,2,5]

[1] *School of Electronic and Optical Engineering, Nanjing University of Science and Technology, No. 200 Xiaolingwei Street, Nanjing, Jiangsu Province 210094, China*

[2] *Jiangsu Key Laboratory of Spectral Imaging and Intelligent Sense, Nanjing University of Science and Technology, Nanjing, Jiangsu Province 210094, China*

[3]*Co-first authors with equal contribution*

[4] *dlzheng@njust.edu.cn*

[5] *eohj@njust.edu.cn*

**Abstract**: Fringe projection profilometry (FPP) has become increasingly important in dynamic 3-D shape measurement. In FPP, it is necessary to retrieve the phase of the measured object before shape profiling. However, traditional phase retrieval techniques often require a large number of fringes, which may generate motion-induced error for dynamic objects. In this paper, a novel phase retrieval technique based on deep learning is proposed, which uses an end-to-end deep convolution neural network to transform a single or two fringes into the phase retrieval required fringes. When the object's surface is located in a restricted depth, the presented network only requires a single fringe as the input, which otherwise requires two fringes in an unrestricted depth. The proposed phase retrieval technique is first theoretically analyzed, and then numerically and experimentally verified on its applicability for dynamic 3-D measurement.

## 1. Introduction

Dynamic three-dimensional (3-D) has been widely used in applications, such as bio-medicine [1], reverse engineering [2], and face recognition [3], etc. A typical structured light technique, fringe projection profilometry (FPP) is often used because of its attributes of high-resolution, high-speed, and high-resolution [4], etc.

FPP first calculates the desired phase by using a phase-shifting algorithm [5–7] or transform based algorithm [8–10]. The phase-shifting algorithm requires at least three phase-shifted sinusoidal fringes, which may generate motion-induced error for dynamic objects [11,12]. The transform-based algorithm can use a single sinusoidal fringe to calculate the desired phase, but it is difficult to preserve the object's edges [13]. Deep learning has been recently introduced to calculate the desired phase, which can directly transform a single fringe into the desired phase, and simultaneously preserve the object's edges [14,15].

The calculated phase requires to be unwrapped to obtain the absolute phase, because it is always wrapped in a range of $(-\pi, \pi]$ [16]. Therefore, a phase unwrapping process is necessary for FPP, which can be classified into two categories of spatial [17] and temporal phase unwrapping [18].

The former often fails for a complex surface because of local error propagation [19]. The latter is commonly used in real measurement, but requires a large number of fringes in forms, such as binary [20], ternary-coded [21], phase-coded [22], or multi-frequency [23], etc.

The multi-frequency method was originated from laser interferometry [18], which can be directly applied to FPP by using two or multiple sets of phase-shifted sinusoidal fringes [19]. In a noise-free system, the way using two sets works well. But in a real noise system, multiple sets are required [24]. Deep learning has been introduced to reduce the required fringes [14], but still requires at least three fringes for successful phase unwrapping. In addition, deep learning has also been introduced to transform a single fringe into 3-D shapes, but generates a non-ignorable measurement error up to 2 mm [25]. Therefore, it has great importance to reduce the fringes, but simultaneously preserve the accuracy, especially for dynamic 3-D measurement.

Deep learning has been widely used in image transformation tasks [26], such as segmentation[27], super-resolution [28], and style transfer [29], etc. In this paper, deep learning is introduced to design a fringe pattern transformation network (FPTNet), which can be trained to transform a single or two fringes into the phase retrieval required fringes, and enables a novel phase retrieval technique detailed as follows:

(i) To calculate the desired phase, FPTNet is trained to transform a single sinusoidal fringe into one set of phase-shifted sinusoidal fringes with the same fringe frequency.

(ii) To obtain the absolute phase, FPTNet is trained to transform a single or two sinusoidal fringes into multiple sets of phase-shifted sinusoidal fringes with different frequencies.

It should be noted that, FPTNet requires a single fringe as the input when the object's surface is located in a restricted depth (i.e., never introducing a phase shift more than one fringe period), which otherwise requires two fringes in an unrestricted depth, as explained in Section 3.2. Both simulations and experiments verify that: the proposed phase retrieval technique can calculate the desired phase accurately, and obtain the absolute phase correctly, which shows great potential for dynamic 3-D measurement.

The rest of the paper is organized as follows. Section 2 illustrates the traditional phase retrieval technique. Section 3 introduces the proposed phase retrieval technique using deep learning. Section 4 gives experiments. Section 5 concludes this paper.

## 2. Traditional phase retrieval technique

To calculate the desired phase, FPP using the phase-shifting algorithm requires one set of phase-shifted sinusoidal fringes as [30]

$$I^f = I_n^f, n = 1, 2, \cdots \quad (1)$$

where $N$ denotes the number of phase steps, $f$ denotes the fringe frequency, and $I_n^f$ denotes the $n$-th fringe in the set of $I^f$, which can be described by

$$I_n^f(x, y) = a(x, y) + b(x, y)\cos[\varphi^f(x, y) + \delta_n], \quad (2)$$

where $(x, y)$ denotes the image coordinate, $a(x, y)$ denotes the background, $b(x, y)$ denotes the modulation, $\delta_n = 2\pi \times (n-1)/N$ denotes the amount of phase shift, and $\varphi^f(x, y)$ denotes the desired phase, which can be calculated by using a least-squares algorithm [31]

$$\varphi^f(x,y) = \tan^{-1} \frac{-\sum_{n=1}^{N} I_n^f(x,y)\sin\delta_n}{\sum_{n=1}^{N} I_n^f(x,y)\cos\delta_n}. \tag{3}$$

To obtain the absolute phase, FPP using the multi-frequency method requires multiple sets of phase-shifted sinusoidal fringes with exponentially increased frequencies as [24]

$$I^{f_i}, i = 1, 2, \cdots \tag{4}$$

where $f_i = 2^{i-1}$ denotes the fringe frequency of the $i$-th set, and $s$ denotes the number of the used sets. For simplicity, we omit the notation $(x,y)$ hereafter. Because $f_1 = 1$, we have $\Phi^{f_1} = \varphi^{f_1}$. Therefore, the absolute phase can be obtained by [32]

$$\Phi^{f_i} = 2\pi K^{f_i} + \varphi^{f_i} = 2\pi INT\left(\frac{2\Phi^{f_{i-1}} - \varphi^{f_i}}{2\pi}\right) + \varphi^{f_i}, i = 2, 3, \cdots s \tag{5}$$

where $\varphi^{f_i}$ denotes the $i$-th phase calculated from $I^{f_i}$, $K^{f_i}$ denotes the corresponding fringe order, and $INT[\bullet]$ gives the nearest integer.

As we know, FPP often uses sinusoidal fringes with a high frequency to achieve a high measurement accuracy [33], and thus $\Phi^{f_s}$ is selected for the 3-D reconstruction by combining with system calibrated parameters [34]. In real measurement, seven or eight sets of sinusoidal fringes are often required to satisfy the demand of the projector's resolution [23].

## 3. The proposed phase retrieval technique based on deep learning

### 3.1 Principle

The proposed phase retrieval technique using deep learning, first transforms a single or two sinusoidal fringes into the phase retrieval required sinusoidal fringes, and then retrieves the phase from these transformed fringes.

A fringe pattern transformation network (FPTNet) is designed for the fringe-to-fringe transformation task [35], which includes two sub-networks, fringe pattern transformation network for the phase calculation (i.e., FPTNet-C), and fringe pattern transformation network for the phase unwrapping (i.e., FPTNet-U).

It should be noted that we divides FPTNet into two sub-networks, which aims to explain it from the two processes of phase calculation and unwrapping clearly. Actually, only a network is required to achieve phase retrieval.

FPTNet-C is trained to transform a single sinusoidal fringe into one set of phase-shifted sinusoidal fringes with the same fringe frequency. In FPTNet-U, when the object's surface does not introduce a phase shift more than one fringe period, FPTNet-U I is trained to transform one sinusoidal fringe into multiple sets of sinusoidal fringes with different frequencies. Otherwise, FPTNet-U II is trained to transform two sinusoidal fringes into multiple sets of sinusoidal fringes with different frequencies.

FPTNet includes two steps of training and testing. In the training step, FPTNet is trained to learn the transformation from the input to the output by minimizing the difference between the outputted fringes and the actual fringes captured by the camera [36]. This difference can be described by a loss function, and the actual fringes are the ground truth of FPTNet. In the testing step, the trained FPTNet directly outputs the desired fringes from the input [37].

Taking FPP using $s$ sets as an example, we detail the proposed phase retrieval technique as follows. FPTNet outputs $s$ sets of phase-shifted sinusoidal fringes with the input of a single or two fringes selected from $I^{f_i}, i=1,2,...,s$. For convenience, the frequency of the outputted fringes is denoted by $f^{out}$, the frequency of the output of the $i$-th set can be denoted by $f_i^{out}$. Therefore, the outputted $s$ sets of fringes can be denoted by $I^{f_i^{out}}, i=1,2,...,s$, where the $i$-th set can be described by $I^{f_i^{out}} = I_n^{f_i^{out}}, n=1,2,...,N$.

For clarity, the diagram of the proposed phase retrieval technique is provided in Fig. 1. FPTNet-C outputs $I_n^{f_s^{out}}, n=1,2...,N$ with the input of $I_1^{f_s}$. By substituting $I_n^{f_s}$ with $I_n^{f_s^{out}}$ in Eq. (3), the desired phase of $\varphi^{f_s^{out}}$ can be calculated.

FPTNet-U I outputs $I^{f_i^{out}}, i=1,2,...,s-1$ with the input of $I_1^{f_s}$, and FPTNet-U II outputs $I^{f_i^{out}}, i=1,2,...,s-1$ with the input of $I_1^{f_s}$ and $I_1^{f_l}$ (Note: $f_l < f_s$). By reusing Eq. (3) $s-1$ times, $\varphi^{f_i^{out}}, i=1,2,\cdots$ can be calculated. Therefore, $\Phi^{f_i^{out}}, i=1,2,\cdots s$ can be obtained by using Eq. (5).

It should be noted that, $I^{f_i^{out}}$ may has a learning difference from $I^{f_i}$. This difference can be descibed as $\Delta I^{f_i^{out}} = I^{f_i^{out}} - I^{f_i}$, and results in the phase error as $\Delta\varphi^{f_i^{out}} = \varphi^{f_i^{out}} - \varphi^{f_i}$, which may generate wrong fringe order as $\Delta K^{f_i^{out}} = K^{f_i^{out}} - K^{f_i}$. According to Eq. (5), $\Delta K^{f_i^{out}}$ can be described by [38]

$$\Delta K^{f_i^{out}} = \frac{2\Delta\varphi^{f_{i-1}^{out}} - \Delta\varphi^{f_i^{out}}}{2\pi}, i=2,3,\cdots \qquad (6)$$

which should be less than 0.5 for a correct phase unwrapping. That is to say, the phase error should satisfy

$$|2\Delta\varphi^{f_{i-1}^{out}} - \Delta\varphi^{f_i^{out}}| < \pi, i=2,3,\cdots \qquad (7)$$

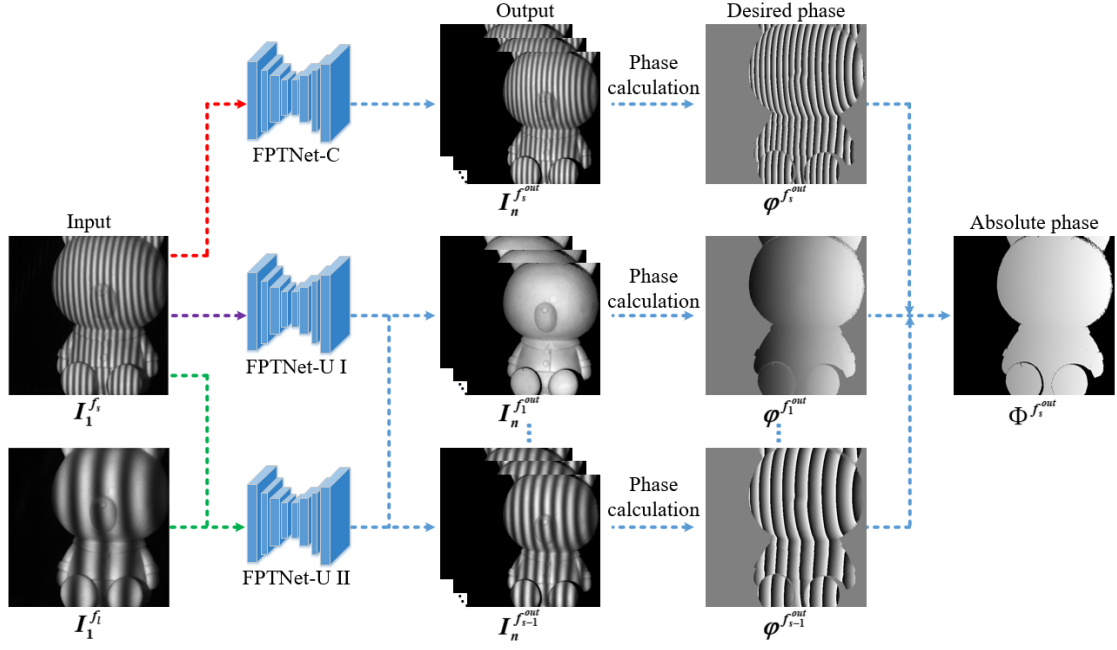

Figure 1: The diagram of the proposed phase retrieval technique using FPTNet.

FPTNet is inspired by Efficient Residual Factorized (ERF) [39, 40], which consists of an encoder and a decoder. Taking FPTNet-C using the input with a size of $512 \times 256$ as an example, the details of structure are provided in Table. 1, where dilated denotes the size of dilated convolution, and output size denotes the size of layer output. In the encoder structure, the input fringe is encoded as the deep semantics by spacial downsampling (Downsample Layer) and the deep features are extracted (ERF Layer). In the decoder, the deep semantics is spacial upsampled and recovered detail features with Upsample Layer and ERF Layers, respectively. Output layer is used to obtain the transformed fringes by the convolution layer.

It should be noted that FPTNet-U has the same setting as FPTNet-C. In FPTNet, we take the outputted fringes with the frequency of $f_i^{out}$ as an example, the loss function can be

$$Loss(\theta_1) = \frac{1}{mN} \sum_{n=1}^{N} \left\| I_n^{f_i^{out}} - I_n^{f_i} \right\|^2, \qquad (8)$$

where $\theta_1$ denotes the parameter space in the network, $m$ denotes the number of pixels and $\|\cdot\|$ is 2-norm.

Table 1. The layer disposal of FPTNet-C. Output sizes are given for an example input of 512x256

|  |  | Type | Output size |
|---|---|---|---|
| **Encoder** | Step 0 | Downsample Layer | 16x512x256 |
|  | Step 1 | Downsample Layer | 64x256x128 |
|  |  | 5 x ERF Layer | 64x256x128 |
|  | Step 2 | Downsample Layer | 128x128x64 |
|  |  | ERF Layer (dilated 2) | 128x128x64 |
|  |  | ERF Layer (dilated 4) | 128x128x64 |
|  |  | ERF Layer (dilated 8) | 128x128x64 |
|  |  | ERF Layer (dilated 16) | 128x128x64 |
| | Step 3 | Repeat Step 2, without Downsample Layer | |
| **Decoder** | Step 4 | Upsample Layer | 64x256x128 |
| | | 2 x ERF Layer | 64x256x128 |
| | Step 5 | Upsample Layer | 16x512x256 |
| | | 2 x ERF Layer | 16x512x256 |
| **Output** | Step 6 | Convolution Layer | 12x512x256 |

*3.2 Type selection strategy for FPTNet-U*

In real, FPTNet-U I and FPTNet-U II can be flexibly selected according to the depth of the object's surface, as illustrated in Fig. 2. Fig. 2(a) gives the FPP system, where $E_{p1} - E_{p2}$ and $E_{c1} - E_{c2}$ represent the optical axis of a projector lens and a camera lens, respectively;, $(x, y, z)$ and $(x^c, y^c)$ denote the world coordinate system and the camera coordinate system, respectively;, $d$ is the distance between the projector system and the camera system, $L$ is the measurement range, $l$ is the distance between the camera system and the reference plane, and the depth of $z = 0$ corresponds to the reference plane.

In the measurement range, five planes are selected with the depths of $z = h_0, p_1, h_1, p_2, h_2$, respectively;, which introduce phase shifts of $\Delta\varphi, 2\pi, 2\pi + \Delta\varphi, 4\pi, 4\pi + \Delta\varphi$ to the reference plane, respectively. The object's surface may be located in any plane due to the object's dynamic movement. Along the camera optical axis, three object's surface points of *A, B* and *C* are selected from the above three planes with the depths of $z = h_0, h_1, h_2$, respectively. It can be assumed that the selected three surface reflect the projected light to the same camera image pixel of $(x_0^C, y_0^C)$.

As shown in Fig. 2(b), we have $\varphi_A(x_0^C, y_0^C) = \varphi_B(x_0^C, y_0^C) = \varphi_C(x_0^C, y_0^C)$ and $\Phi_A(x_0^C, y_0^C) = \Phi_B(x_0^C, y_0^C) - 2\pi = \Phi_C(x_0^C, y_0^C) - 4\pi$. It is apparent that: (i) when the object's surface is located in the depth range of $(0, p_1)$, the calculated phase of each pixel corresponds to a unique absolute phase; (ii) when the object's surface is located in the depth range of $(p_1, L)$, the calculated phase of any pixel may correspond to multiple absolute phases. When one pixel corresponds to multiple absolute phases, FPTNet-U I cannot differentiate them, and may lead to a wrong fringe transformation, where FPTNet-U II is required.

As aforementioned, the selection strategy for FPTNet-U is given as follows:

(i) When the object's surface does not introduce a phase shift more than $2\pi$ (i.e., one fringe period), FPTNet-U I is selected with the input of $I_1^{f_s}$.

(ii) When the object's surface introduces a phase shift more than $2\pi$, FPTNet-U II is selected with the input of $I_1^{f_s}$ and $I_1^{f_l}$.

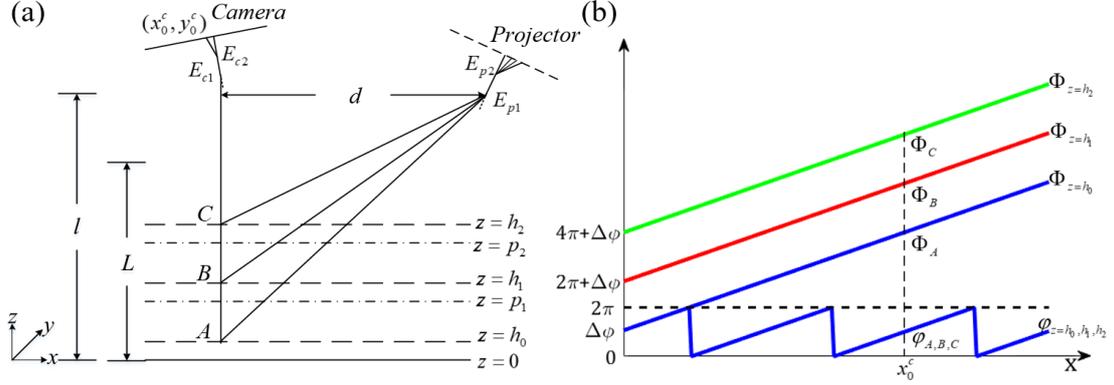

Figure 2: The diagram of type selection strategy for FPTNet-U. (a) The FPP system with three object's surface points of $A$, $B$ and $C$. (b) The wrapped phase and absolute phase corresponding to points of $A$, $B$ and $C$.

The highest frequency of $f_s$ can be selected according to the projector resolution [41]. According to the following simulation, and experiments provided in Section 4, the lower frequency of $f_l$ can be selected as follows:

(i) $f_l$ can be selected as any frequency that does not a common divisor of $f_s$, e.g., $f_s = 64, f_l = 45$.

(ii) $f_l$ can also be selected as a common divisor of $f_s$, when the object's surface does not introduce a phase shift more than $2\pi$, e.g., $f_s = 64, f_l = 16$.

It should be noted that, when vertical fringe patterns are projected, the restricted depth range of $z_{th}$ can be determined by [42]

$$z_{th} = \frac{lc_1}{f_s d}, \qquad (13)$$

where $c_1$ is the projected length of the projector in the horizontal direction.

A simulation is provided to verify the above selection strategy. For simplicity, both the object and the reference plane are assumed to have the same reflectivity. The object's surface is generated as follows: first, a random matrix is generated by randomly selecting a amplitude from the range of (5, 55); second, a Gaussian filter with the standard deviation range of $(2,125)$ is used to filter the generated random matrix; third, the filtered result is scaled and moved to generate the desired object's surface of $z(x,y)$; finally, according to Eq. (2), one set of fringes of $I_n^f$ with a frequency of $f$ can be generated by [42, 43]

$$I_n^f(x,y) = 127.5 + 127.5\cos[z(x,y)fc + \delta_n], n = 1, 2, \cdots \quad (10)$$

where $N$ denotes the number of phase steps, and $c$ is a constant determined by the measurement system.

Dataset I and dataset II are generated for FPTNet-U I and FPTNet-U II, respectively;, which are deformed by objects' surfaces introducing phase shifts less and more than $2\pi$, respectively. According to Eq. (10), we select $c = 1/35$ and $f_s = 35$, and thus we have $z_{th}$=6.28. Therefore, dataset I and dataset II are constructed when objects' surfaces are located in ranges of $[0, 6.28]$ and $[0, 25]$, respectively. FPTNet-U I selects the input with the frequency of $f_s = 35$, and FPTNet-U II selects the input with two frequencies of $f_s = 35$ and $f_l = 30$. Both FPTNet-U I and FPTNet-U II output fringes with the frequency of $f^{out} = 25$. The twelve-step phase-shifting algorithm is selected to calculate the desired phase.

For each dataset, the training set, validation set and testing set are generated from 500 surfaces, 100 surfaces and 50 surfaces, respectively. The resolution of these generated fringes is selected as $300 \times 300$. For each surface, two sets with two frequencies of 35 and 30 are generated, respectively;, and another set with the frequency of 25 is also generated as the ground-truth. Therefore, the training set, validation set and testing set include 18000, 3600 and 1800 fringes, respectively. The validation set is aimed to monitor during training the accuracy of the neural network on the data [14].

The results of FPTNet-U I are shown in Fig. 3. One surface as shown in Fig. 3(a) is selected. The corresponding input and output are shown in Fig. 3(b) and 3(c), respectively. The actual phase and the desired phase are calculated by using Eq. (3) twice. By subtracting the actual phase from the desired phase, the phase error is obtained and shown in Fig .3(d). The phase error is generally less than 0.01 rad, which is much less than the threshold of $\pi$ given in Eq. (7). FPTNet-U I can obviously obtain the desired phase for a correct phase unwrapping.

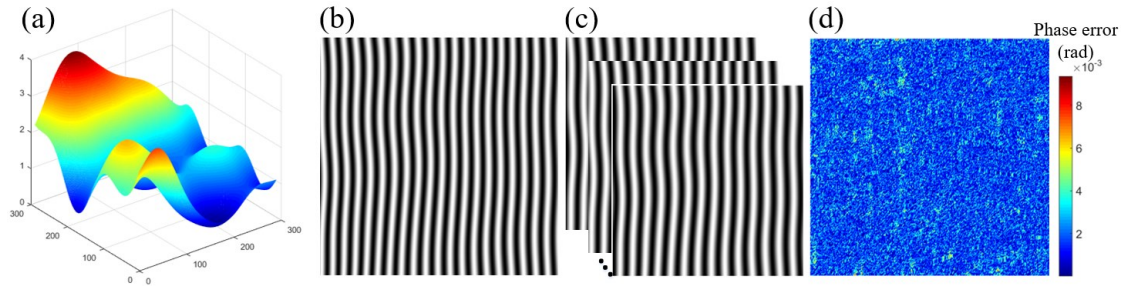

Figure 3: The results of dataset I for FPTNet-U I. (a) The surface. (b) The input. (c) The output. (d) The phase error.

To verify FPTNet-U II, one surface as shown in Fig. 4(a) is selected, and two fringes with two frequencies of 35 and 30 are selected as the input and shown in Fig. 4(b). FPTNet-U I is also tested on the same surface for the comparison, but only selects the single fringe with the frequency of 35 as the input. The results of FPTNet-U I and FPTNet-U II are provided in Fig. 4(c) and 4(d), respectively, where the left, middle, and right show the outputted fringes, the grayscale error of the 12-th outputted fringe, and the phase error, respectively. As we can see, FPTNet-U I outputs fringes with a large grayscale error, and results in a calculated phase with a phase error larger than $\pi$, which will lead to a wrong phase unwrapping according to Eq. (7). FPTNet-U II outputs fringes with a small grayscale error, and also results in a phase with a small phase error less than 0.02rad, which will enable a correct phase unwrapping. Both FPTNet-U I and FPTNet-U II are also tested on datasets with other frequencies, and similar results can also be obtained.

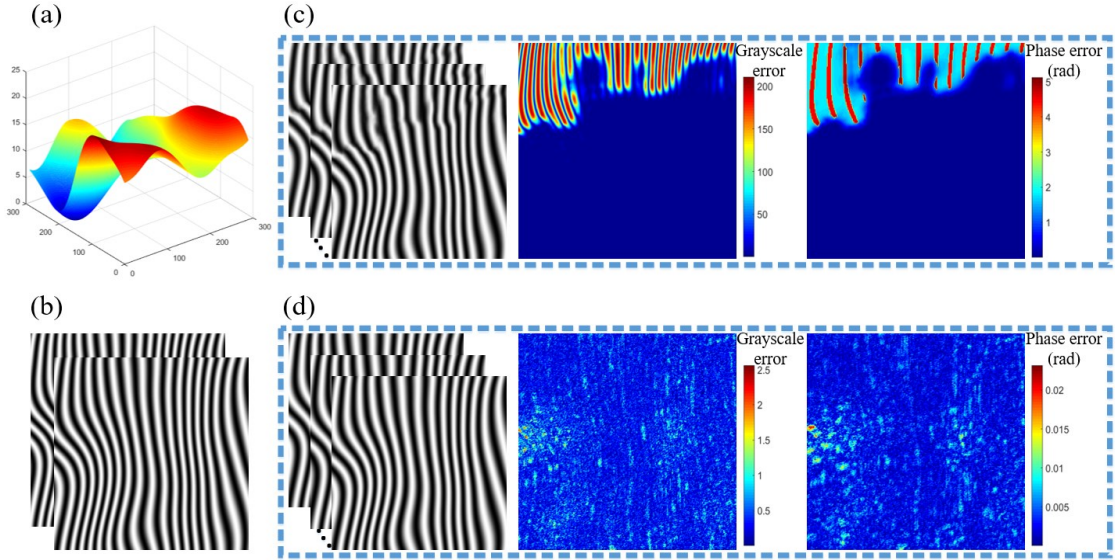

Figure 4: The results of dataset II for FPTNet-U. (a) The surface. (b) The input. (c) The results of FPTNet-U I. (d) The results of FPTNet-U I. For (c) and (d), each column shows the outputted fringe, the grayscale error, and the phase error.

## 4. Experiment

We first verify the proposed phase retrieval technique, and then provide real dynamic 3-D measurements to evaluate its performance. The experiment system includes a DLP6500 projector with a resolution of $1920 \times 1080$, a Basler acA800-510um CMOS camera with a resolution of $496 \times 496$, and a Computar lens of 12mm focal length. The distance between the camera system and the reference plane, the projected length of the projector in the horizontal direction, the distance between the projector system and the camera system are about $l = 1000mm, c_1 = 280mm$, and $d = 60mm$, respectively.

To verify the proposed phase retrieval technique, a twelve-step ($N = 12$) phase-shifting algorithm is also used to calculate the desired phase, and a multi-frequency ($s = 7$, $f_i = 2^{i-1}, i = 1, 2, ..., 7$) phase unwrapping method is used to obtain the absolute phase. Therefore, for each frequency, the outputted set should be $I^{f_i^{out}} = I_n^{f_i^{out}}, n = 1, 2, ...$ . For comparison, another three sets are also outputted with three frequencies of 45, 90 and 120, respectively.

Two datasets of I and II are constructed in a restricted and an unrestricted depths, respectively. In our system, the restricted depth of $z_{th}$ and the measurement range are about $73mm$ and 250mm, respectively. Therefore, dataset I is constructed when the measured object is located in the range of [0, 73mm]. The training set, validation set and testing set of dataset I contain 50 scenes, 20 scenes and 20 scenes, respectively. Dataset II is constructed when the measured object is located in the range of [0, 250mm]. The training set, validation set and testing set of dataset II contain 150 scenes, 25 scenes and 25 scenes, respectively. For each scene, ten sets are captured as the ground truth.

### 4.1 Experiments for FPTNet-C

On both dataset I and dataset II, FPTNet-C outputs $I^{f_7^{out}}$ with the input of $I_1^{f_7}$. By using Eq. (3), the desired phase of $\varphi^{f_7^{out}}$ and the actual phase of $\varphi^{f_7}$ can be calculated from $I^{f_7^{out}}$ and $I^{f_7}$, respectively.

The experimental results are provided in Fig. 5 and Fig. 6. In Fig. 5, the left, middle and right box show the input, the output and the desired phase, respectively. By subtracting the actual phase from the desired phase, the phase error of FPTNet-C is obtained and shown in Fig. 6(a). The mean value of the phase error is 0.05 rad, which is comparable with 0.02rad of the random phase error obtained by using the traditional phase-shifting algorithm [44]. In Fig. 6(a), two areas in Fig. 6(a) containing complex and smooth textures are enlarged and reshown in Fig. 6(b) and 6(c), respectively. It is apparent that, the phase error can be up to 0.6 rad in the complex surface, but it is generally less than 0.07 rad in the smooth area.

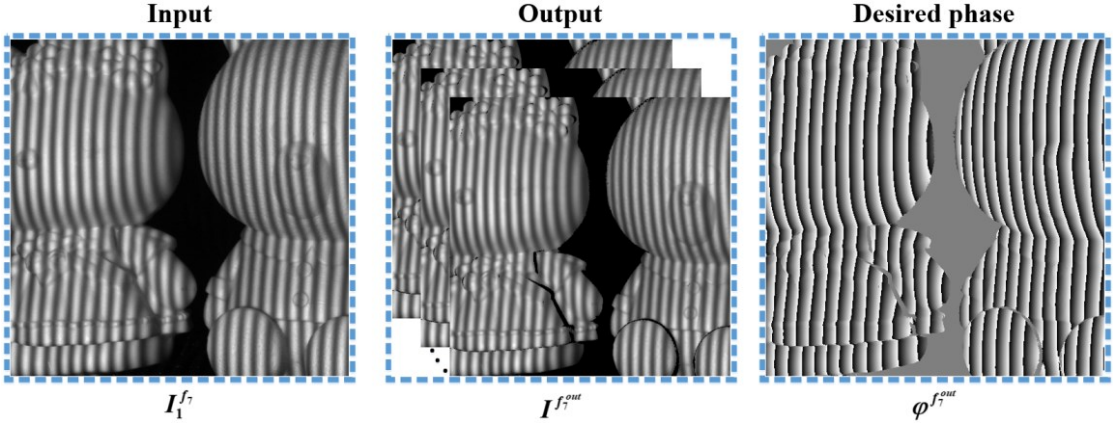

Figure 5: The results of dataset II for FPTNet-C, and the box of each column shows the input, the output and the desired phase.

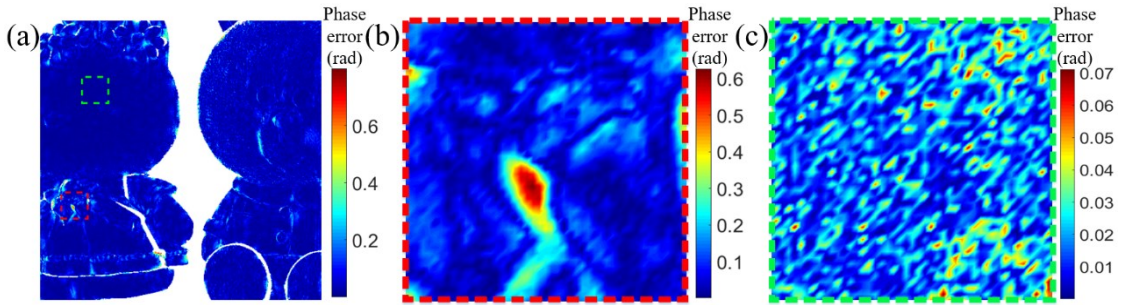

Figure 6: (a) The phase error of FPTNet-C. (b) The corresponding enlarged detail of the red box in (a). (c) The corresponding enlarged detail of the green box in (a).

FPTNet-C is also verified on other different frequencies selected from the range of [4, 120]. The mean values of the phase error range from 0.04rad to 0.06rad, and FPTNet-C performs constantly well according to different frequencies, as shown in Fig. 7. For clarity, the results of the three frequencies of 4, 32 and 120 are provided in Fig. 7(a), 7(b) and 7(c), respectively;, and each row shows the input, the output, the desired phase, and the phase error.

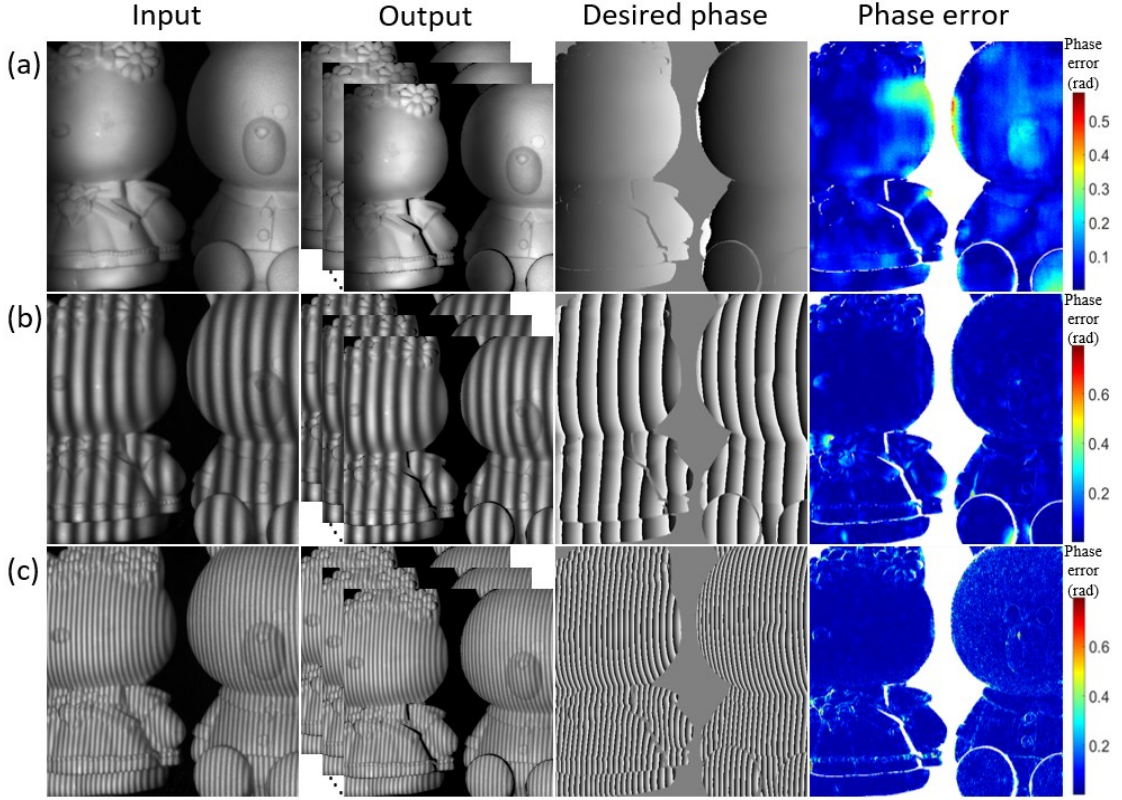

Figure 7: (a) The results with the frequency of 4. (b) The results with the frequency of 32. (c) The results with the frequency of 120. Each row shows the input, the output, the desired phase and the phase error.

FPTNet-C is then verified on phase-shifting algorithms with different phase steps selected from the range of [3, 12], and the fringe of $I_1^{f_j}$ is also selected as the input. The mean value of the phase error is nearly constant at around 0.04rad. FPTNet-C also performs constantly well for phase-shifting algorithms with different phase steps.

### 4.2 Experiments for *FPTNet-U*

FPTNet-U I is verified on dataset I. FPTNet-U I outputs $I^{f_i^{out}}, i=1,2,...,6$ with the input of $I_1^{f_j}$. Therefore, six desired phases of $\varphi^{f_i^{out}}, i=1,2,...,6$ are calculated from these outputted six sets of fringes. Seven absolute phases of $\Phi^{f_i^{out}}, i=1,2,...,7$ can be obtained from $\varphi^{f_i^{out}}, i=1,2,...,6$ and the above FPTNet-C obtained $\varphi^{f_j^{out}}$. For comparison, $\Phi^{f_i}, i=1,2,...,6$ can be obtained from the actual phase of $\varphi^{f_i}, i=1,2,...,6$.

The results of FPTNet-U I are provided in Fig. 8, where the left, middle and right box show the input, the output and the absolute phase, respectively. The desired phases with different frequencies can be calculated by the same way as FPTNet-C, which also have similar accuracy as FPTNet-C. Therefore, FPTNet-U I can obtain the same correct fringe order for the phase unwrapping as the traditional multi-frequency method, which will enable a correct phase unwrapping.

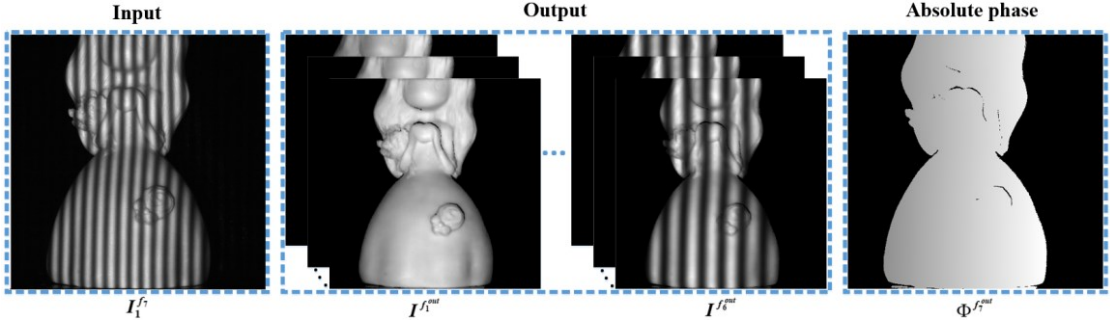

Figure 8: The results of dataset I for FPTNet-U I, and the box of each column shows the input, the output and the absolute phase.

FPTNet-U I is also verified on dataset II, as shown in Fig. 9. We take the outputted fringes with the frequency of $f_5^{out}=32$ as an example. For comparison, $I_3^{f_5}$ is selected as shown in Fig. 9(a), and the corresponding outputted $I_3^{f_5^{out}}$ is shown 9(b). The grayscale error between $I_3^{f_5}$ and $I_3^{f_5^{out}}$ is shown in Fig. 9(c), which generates a phase error larger than $\pi$ as shown in Fig. 9(d). According to Eq. (7), the calculated phase will lead a wrong phase unwrapping.

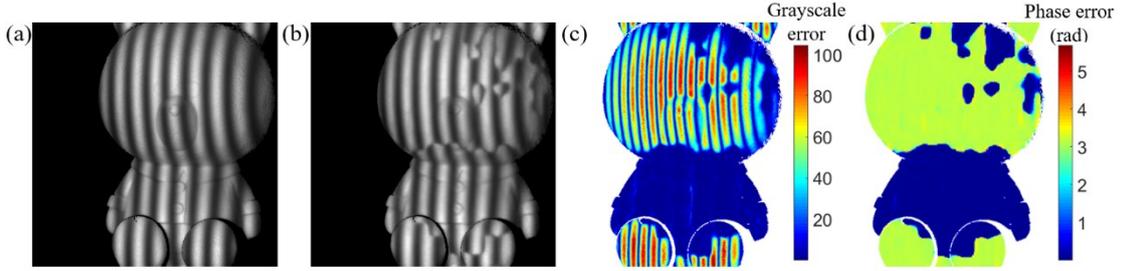

Figure 9: The object located in an unrestricted depth. (a) The ground-truth. (b) The output of FPTNet-U I. (c) The grayscale error. (d) The phase error.

In this condition, FPTNet-U II is necessary for dateset II. FPTNet-U II has the same output as FPTNet-U I, but selects the input of $I_1^{f_7}$ and $I_1^{f_5}$. The results of FPTNet-U II are provided in Fig. 10, where the left, middle and right box show the input, the output and the absolute phase, respectively. The desired phases calculated from FPTNet-U II transformed fringes also have similar accuracy as FPTNet-C, which can also enable a correct phase unwrapping.

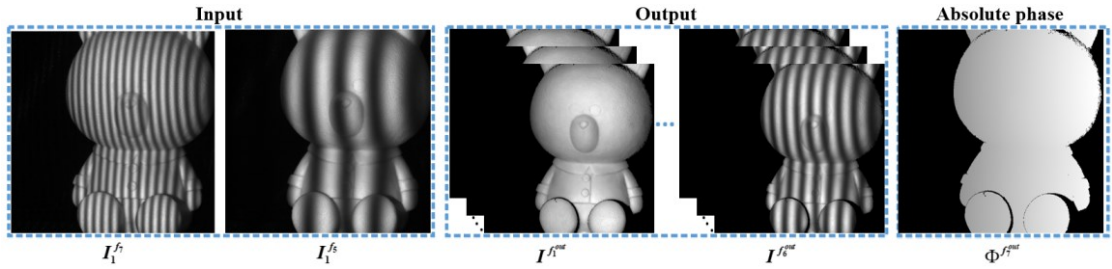

Figure 10: The results of datasets II for FPTNet-U II, and the box of each column shows the input, the output and the absolute phase.

### 4.3 Real dynamic experiments

In a restricted depth, an electric fan in rotation is measured. FPTNet outputs $I_i^{f_i^{out}}, i=1,2,...,7$ with the input of $I_1^{f_7}$. Therefore, seven desired phases of $\varphi^{f_i^{out}}, i=1,2,...,7$ are calculated, and $\Phi^{f_7^{out}}$ can be obtained for the 3-D reconstruction.

The measurement results are provided in Fig. 11. Four inputs with each including a single fringe of $I_1^{f_7}$ are selected, and their corresponding 3-D shapes are shown in Fig. 11(a) and 11(b), respectively. For clarity, the reconstructed 3-D shapes of the dynamic fan are provided in Visualization 1. The dynamic 3-D object located in a restricted depth can be measured without motion-induced errors [45].

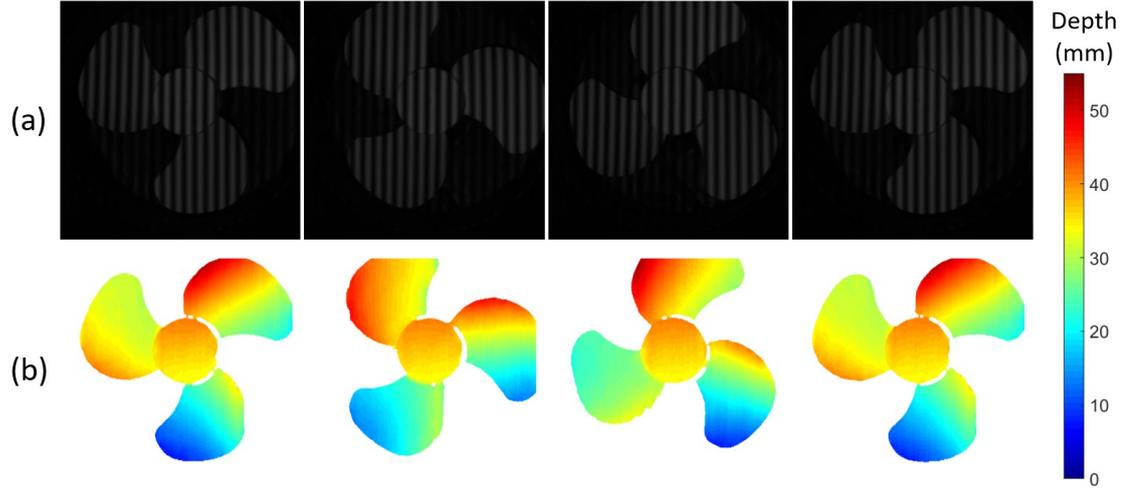

Figure 11: A rotating electric fan in a restricted depth. (a) The input. (b) The corresponding 3-D shapes (Visualization 1).

In an unrestricted depth, a toy with dynamic movement is measured. The measurement results are provided in Fig. 12. Four inputs with each including a single fringe of $I_1^{f_7}$ are selected, and their corresponding 3-D shapes are shown in Fig. 12(a) and 12(c), respectively. Another four inputs are selected with each including two fringes of $I_1^{f_7}$ and $I_1^{f_5}$, which are shown in Fig. 12(a) and 12(b), respectively. Their corresponding 3-D shapes are shown in Fig. 12(d). For clarity, the reconstructed 3-D shapes of the moving toy are provided in Visualization 2, where results I and results II represent the results reconstructed from a single or two fringes as the input, respectively. The dynamic 3-D object located in an unrestricted depth can also be measured.

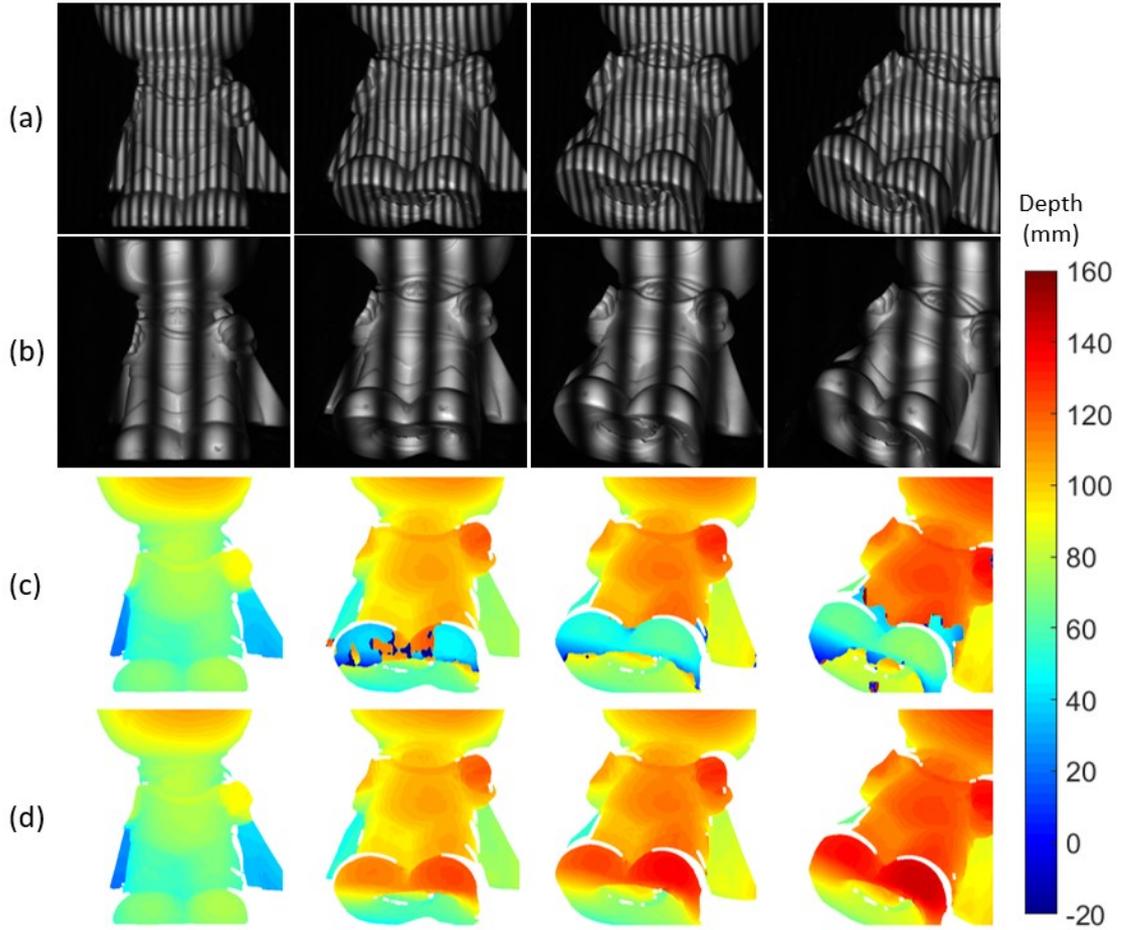

Figure 12: A toy moving in an unrestricted depth. (a) The input of $I_1^{f_1}$. (b) The input of $I_1^{f_5}$. (c) The corresponding 3-D shapes with a single fringe as the input. (d) The corresponding 3-D shapes with two fringes as the input (Visualization 2).

## 4  Conclusion

In this paper, a fringe pattern transformation network (FPTNet) is designed for the fringe-to-fringe transformation task using deep learning, which can be trained to transform a single or two sinusoidal fringes into multiple sets of phase-shifted sinusoidal fringes. A single or two fringes can be flexibly selected as the input of FPTNet according to the object located depth. Based on the FPTNet transformed fringes, the desired phase can be accurately calculated, and the absolute phase can be correctly obtained, which can reconstruct 3-D shapes without motion-induced errors for dynamic objects.

## References

1. Y. Wang, J. I. Laughner, I. R. Efimov, and S. Zhang, "3d absolute shape measurement of live rabbit hearts with a superfast two-frequency phase-shifting technique," Opt. Express **21**, 5822–5832 (2013).

2. B. Li and S. Zhang, "Novel method for measuring a dense 3d strain map of robotic flapping wings," Meas. Sci. Technol. **29**, 045402 (2018).

3. J. Xue, Q. Zhang, C. Li, W. Lang, M. Wang, and Y. Hu, "3d face profilometry based on galvanometer scanner with infrared fringe projection in high speed," Appl. Sci. **9** (2019).

4. Z. Wang, D. A. Nguyen, and J. C. Barnes, "Some practical considerations in fringe projection profilometry,"

Opt.Lasers Eng. **48**, 218–225 (2010).

5. S. Zhang and S.-T. Yau, "High-resolution, real-time 3d absolute coordinate measurement based on a phase-shifting method," Opt. Express **14**, 2644–2649 (2006).

6. T. Tao, Q. Chen, J. Da, S. Feng, Y. Hu, and C. Zuo, "Real-time 3-d shape measurement with composite phase-shifting fringes and multi-view system," Opt. express **24**, 20253–20269 (2016).

7. B. Pan, Q. Kemao, L. Huang, and A. Asundi, "Phase error analysis and compensation for nonsinusoidal waveforms in phase-shifting digital fringe projection profilometry," Opt. Lett. **34**, 416–418 (2009).

8. Q. Kemao, "Windowed fourier transform for fringe pattern analysis," Appl. Opt. **43**, 2695–2702 (2004).

9. X. Su and Q. Zhang, "Dynamic 3-d shape measurement method: A review," Opt. Lasers Eng. **48**, 191 – 204 (2010). Fringe Projection Techniques.

10. J. Zhong and J. Weng, "Spatial carrier-fringe pattern analysis by means of wavelet transform: wavelet transform profilometry," Appl. optics **43**, 4993–4998 (2004).

11. L. Lu, Z. Jia, Y. Luan, and J. Xi, "Reconstruction of isolated moving objects with high 3d frame rate based on phase shifting profilometry," Opt. Commun. **438**, 61 – 66 (2019).

12. S. Feng, C. Zuo, T. Tao, Y. Hu, M. Zhang, Q. Chen, and G. Gu, "Robust dynamic 3-d measurements with motion-compensated phase-shifting profilometry," Opt. Lasers Eng. **103**, 127–138 (2018).

13. Q. Kemao, "Two-dimensional windowed fourier transform for fringe pattern analysis: Principles, applications and implementations," Opt. Lasers Eng. **45**, 304 – 317 (2007). Phase Measurement Techniques and their applications.

14. S. Feng, C. Zuo, W. Yin, G. Gu, and Q. Chen, "Micro deep learning profilometry for high-speed 3d surface imaging," Opt. Lasers Eng. **121**, 416 – 427 (2019).

15. S. Feng, Q. Chen, G. Gu, T. Tao, L. Zhang, Y. Hu, W. Yin, and C. Zuo, "Fringe pattern analysis using deep learning," Adv. Photonics **1**, 1 – 7 (2019).

16. W. Yin, C. Zuo, S. Feng, T. Tao, Y. Hu, L. Huang, J. Ma, and Q. Chen, "High-speed three-dimensional shape measurement using geometry-constraint-based number-theoretical phase unwrapping," Opt. Lasers Eng. **115**, 21 – 31 (2019).

17. M. Takeda and K. Mutoh, "Fourier transform profilometry for the automatic measurement of 3-d object shapes," Appl. optics **22**, 3977–3982 (1983).

18. S. Zhang, "Absolute phase retrieval methods for digital fringe projection profilometry: A review," Opt. Lasers Eng. **107**, 28 – 37 (2018).

19. C. Zuo, S. Feng, L. Huang, T. Tao, W. Yin, and Q. Chen, "Phase shifting algorithms for fringe projection profilometry: A review," Opt. Lasers Eng. **109**, 23 – 59 (2018).

20. G. Sansoni, M. Carocci, and R. Rodella, "Three-dimensional vision based on a combination of gray-code and phase-shift light projection: analysis and compensation of the systematic errors," Appl. optics **38**, 6565–6573 (1999).

21. D. Zheng, Q. Kemao, F. Da, and H. S. Seah, "Ternary gray code-based phase unwrapping for 3d measurement using

binary patterns with projector defocusing," Appl. optics **56**, 3660–3665 (2017).

22. Y. Wang and S. Zhang, "Novel phase-coding method for absolute phase retrieval," Opt. letters **37**, 2067–2069 (2012).

23. C. Zuo, L. Huang, M. Zhang, Q. Chen, and A. Asundi, "Temporal phase unwrapping algorithms for fringe projection profilometry: A comparative review," Opt. Lasers Eng. **85**, 84 – 103 (2016).

24. J. M. Huntley and H. O. Saldner, "Shape measurement by temporal phase unwrapping: comparison of unwrapping algorithms," Meas. Sci. Technol. **8**, 986–992 (1997).

25. H. Nguyen, H. Li, Q. Qiu, Y. Wang, and Z. Wang, "Single-shot 3d shape reconstruction using deep convolutional neural networks," arXiv preprint arXiv:1909.07766 (2019).

26. K. Jia, X. Wang, and X. Tang, "Image transformation based on learning dictionaries across image spaces," IEEE Transactions on Pattern Analysis Mach. Intell. **35**, 367–380 (2013).

27. E. Shelhamer, J. Long, and T. Darrell, "Fully convolutional networks for semantic segmentation," IEEE Transactions on Pattern Analysis Mach. Intell. **39**, 640–651 (2017).

28. C. Dong, C. C. Loy, K. He, and X. Tang, "Image super-resolution using deep convolutional networks," IEEE Transactions on Pattern Analysis Mach. Intell. **38**, 295–307 (2016).

29. J. Johnson, A. Alahi, and L. Fei-Fei, "Perceptual losses for real-time style transfer and super-resolution," in *Computer Vision – ECCV 2016,* B. Leibe, J. Matas, N. Sebe, and M. Welling, eds. (Springer International Publishing, Cham, 2016), pp. 694–711.

30. S. Zhang, "Recent progresses on real-time 3d shape measurement using digital fringe projection techniques," Opt. Lasers Eng. **48**, 149 – 158 (2010). Fringe Projection Techniques.

31. D. Zheng, F. Da, Q. Kemao, and H. S. Seah, "Phase-shifting profilometry combined with gray-code patterns projection: unwrapping error removal by an adaptive median filter," Opt. express **25**, 4700–4713 (2017).

32. S. Zhang, "Phase unwrapping error reduction framework for a multiple-wavelength phase-shifting algorithm," Opt. Eng. **48**, 105601 (2009).

33. D. Zheng and F. Da, "Phase coding method for absolute phase retrieval with a large number of codewords," Opt. Express **20**, 24139–24150 (2012).

34. K. Liu, Y. Wang, D. L. Lau, Q. Hao, and L. G. Hassebrook, "Dual-frequency pattern scheme for high-speed 3-d shape measurement," Opt. Express **18**, 5229–5244 (2010).

35. C. Wang, C. Xu, C. Wang, and D. Tao, "Perceptual adversarial networks for image-to-image transformation," IEEE Transactions on Image Process. **27**, 4066–4079 (2018).

36. K. Yan, Y. Yu, C. Huang, L. Sui, K. Qian, and A. Asundi, "Fringe pattern denoising based on deep learning," Opt. Commun. **437**, 148–152 (2019).

37. J. Shi, X. Zhu, H. Wang, L. Song, and Q. Guo, "Label enhanced and patch based deep learning for phase retrieval from single frame fringe pattern in fringe projection 3d measurement," Opt. express **27**, 28929–28943 (2019).

38. C. Zhang, H. Zhao, and K. Jiang, "Fringe-period selection for a multifrequency fringe-projection phase unwrapping method," Meas. Sci. Technol. **27**, 085204 (2016).

39. E. Romera, J. M. Álvarez, L. M. Bergasa, and R. Arroyo, "Efficient convnet for real-time semantic segmentation," in *2017 IEEE Intelligent Vehicles Symposium (IV),* (2017), pp. 1789–1794.

40. E. Romera, J. M. Álvarez, L. M. Bergasa, and R. Arroyo, "Erfnet: Efficient residual factorized convnet for real-time semantic segmentation," IEEE Transactions on Intell. Transp. Syst. **19**, 263–272 (2018).

41. C. Zuo, T. Tao, S. Feng, L. Huang, A. Asundi, and Q. Chen, "Micro fourier transform profilometry (′ftp): 3d shape measurement at 10,000 frames per second," Opt. Lasers Eng. **102**, 70 – 91 (2018).

42. Y. An, J.-S. Hyun, and S. Zhang, "Pixel-wise absolute phase unwrapping using geometric constraints of structured light system," Opt. Express **24**, 18445–18459 (2016).

43. X. Su, W. Chen, Q. Zhang, and Y. Chao, "Dynamic 3-d shape measurement method based on ftp," Opt. Lasers Eng. **36**, 49–64 (2001).

44. D. Zheng, F. Da, Q. Kemao, and H. S. Seah, "Phase error analysis and compensation for phase shifting profilometry with projector defocusing," Appl. Opt. **55**, 5721–5728 (2016).

45. X. Liu, T. Tao, Y. Wan, and J. Kofman, "Real-time motion-induced-error compensation in 3d surface-shape measurement," Opt. express **27**, 25265–25279 (2019).